\documentclass[aps,twocolumn,groupedaddress,showpacs]{revtex4}

\usepackage{amsmath} 
  \newcommand{\beqa}{\begin{eqnarray}}
  \newcommand{\eeqa}{\end{eqnarray}}
  \newcommand*{\cH}{\mathcal{H}}
  \newcommand*{\hp}{\hat{p}}
  \newcommand*{\hx}{\hat{x}}

  \newcommand*{\da}{\dagger}

  \newcommand*{\la}{\lambda}
    
  \newcommand*{\Eq}[1]{Eq.~(\ref{eq:#1})}
   \newcommand*{\Then}{\Rightarrow}
  \newcommand*{\eq}[1]{(\ref{eq:#1})}

\newcommand*{\ket}[1]{|#1\rangle}

\newcommand*{\bracket}[1]{\langle#1\rangle}

\begin{document}

\title{Defense of ``Impossibility of distant indirect measurement of the quantum Zeno
effect''}

\author{Masanao Ozawa}
\email[]{ozawa@math.is.tohoku.ac.jp}
%\thetaanks{}
\affiliation{Graduate School of Information Sciences,
T\^{o}hoku University, Aoba-ku, Sendai,  980-8579, Japan}

%\date{\today}
%%%%% ABSTRACT %%%%%% insert abstract here
\begin{abstract}
Recently, Wallentowitz and Toschek
[Phys. Rev. A 69, 046101 (2005)]
criticized the assertion made by
Hotta and Morikawa  
[Phys. Rev. A 69, 052114 (2004)]   
that distant indirect measurements do not
cause the quantum Zeno effect,
and claimed that their proof is faulty 
and that their claim is unfounded.
Here, it is shown that the argument given by
Wallentowitz and Toschek includes a
mathematical flaw and that their criticism
is unfounded.\end{abstract}

%%%%% PACS NUMBERS %%%%%

\maketitle
	
% TEXT:

In Ref.~\cite{HM04},
Hotta and Morikawa (HM) claimed that quantum Zeno effect does not
take place in repeated projective measurements of the subspace that is 
preserved under the advanced unitary time evolution.
However, Wallentowitz and Toschek (WT) \cite{WT05} made a criticism
that their proof is faulty so that the above claim is unfounded.
Here, it is shown that the argument given by Wallentowitz and Toschek 
is mathematically flawed so that their criticism is unfounded.

HM considered a system described by a Hilbert space $\cH_{Z}$
with unitary time evolution $U(t)$.   
They called a subspace $\cH_{C}$ of $\cH_{Z}$ a {\em core-zone} subspace 
if we have
the orthogonal decomposition $\cH_{Z}=\cH_{C}\oplus\cH_{W}$ satisfying

(I) any vector $\ket{W}\in\cH_{W}$ satisfies $U(t)\ket{W}\in\cH_{W}$ for
all $t>0$.

The subspace $\cH_{W}$ is called the {\em wave-zone} subspace.
Let $P_C$ and $P_W$ be the projections onto the core-zone subspace and
the wave-zone subspace, respectively.  
HM proved that the survival probability
$
s(t)=|\bracket{e|U(t)|e}|^2
$
of any core-zone state $\ket{e}\in\cH_C$
is not affected by $N$-time projective measurements of $P_C$
carried out at times $t_j$ with $0<t_1<\cdots<t_N<t$.
HM also claimed that this holds despite that 
the core-zone state $\ket{e}\in\cH_C$ may decay into a wave-zone 
state with a positive probability 
$\|P_W U(t)\ket{e}\|^2>0$; 
see Eqs. (19)--(21) of Ref.~\cite{HM04}.

However, WT claimed that condition (I) implies 

(II) $P_W U(t) P_C=0$ for all $t>0$.

This implies that the core-zone state never decays 
into a wave-zone state.
From the above, WT concluded that HM's clam is rather 
trivial.  Here, we shall show that this claim of WT is faulty.

WT's argument runs as follows.
WT considered the Hamiltonian $H(t)$ generating $U(t)$ 
and we concentrate on the time independent case where 
$H(t)=H$ to show their argument is faulty even in this simpler
case.  Then, we have 
\beqa\label{eq:3}
U(t)=e^{-itH/\hbar},
\eeqa 
which WT then expanded in a series of powers of the Hamiltonian as
\beqa\label{eq:4}
U(t)=\sum_{n=0}^{\infty}\frac{1}{n!}(\frac{tH}{i\hbar})^n. 
\eeqa
In the corresponding equation in WT's  argument,
the factor ${1}/{n!}$ is missing; see  \Eq{4} of Ref.~\cite{WT05}.

Then, using the expansion \eq{4} WT claimed that from condition (I) 
one must have

(III) $H\ket{W}\in\cH_{W}$ for any $\ket{W}\in\cH_{W}$.

Then, WT claimed that from condition (III), one can derive the following 
condition, as the relation,
being absent in Ref.~\cite{HM04}, which is the central ingredient that allows WT's
conclusion. 

(IV) $U(t)^{\da}\ket{W}\in\cH_{W}$ for all $t>0$ 
and any $\ket{W}\in\cH_{W}$.

From condition (IV), condition (II) can be derived easily.

We shall show that the above argument along with the logical sequence
(I) $\Then$ (III) $\Then $ (IV) $\Then$ (II)
is faulty.

First of all, it is a well-known mathematical fact that condition (I) does not
imply condition (II).
A counter example is given by $U(t)=e^{-it\hp/\hbar}$ 
with the momentum operator
$\hp$  associated with the coordinate operator $\hx$.
Let $\cH_Z$ be the $L^2$-space of wave functions on the real line.
In this case, it is well-know that $U(t)$ represents the translation 
of wave packets from left to right 
so that $[U(t)f](x)=f(x-t)$ holds for every $f\in\cH_Z$.
Let $\cH_C$ and $\cH_W$ be the spaces of wave functions 
supported in the negative and positive
parts of the coordinate, respectively.
Then, it is quite obvious that every $\ket{W}\in\cH_W$ remains in 
$\cH_W$ but for any $t>0$ there is some $\ket{C}\in\cH_C$ that enters 
$\cH_W$ in time $t$, and thus we conclude that condition (I) holds but
condition (II) does not hold.

Thus,  TW's argument is faulty.
The flaw arises from illegitimate manipulations of unbounded operators.
We should note that condition (III) above is not properly formulated or it is simply
false, since $\ket{W}\in\cH_{W}$ may be outside of the domain of $H$.
Thus, we should reformulate it as 

(III-A)  $H\ket{W}\in\cH_{W}$ holds if $\ket{W}\in \cH_{W}$,
as long as $\ket{W}$ is in the domain of $H$.

WT suggested that (III-A) is obtained from the expansion \eq{4}; however,
as discussed later, the expansion \eq{4} cannot be applied to any $\ket{W}$ 
in the domain of $H$.  Instead of using \Eq{4}, we obtain (III-A) from
Stone's theorem \cite{RS80}, which states that 
\beqa
i\hbar\lim_{t\to 0}\frac{U(t)\ket{W}-\ket{W}}{t}=H\ket{W}
\eeqa
holds for any $\ket{W}$ in the domain of $H$.  Taking the limit for $t>0$,
we conclude that condition (I) implies condition (III-A).

Now,
we should notice that \Eq{3} holds everywhere but \Eq{4} holds
only a dense subset consisting of {\em analytic vectors}; see p.~276 of 
Ref.~\cite{RS80} and p.~200 of Ref.~\cite{RS75}.
In fact, the right hand side of \Eq{3} is defined through the spectral family
$\{E_{\la}\}$ of the self-adjoint operator $H$ as
\beqa
e^{-itH/\hbar}=\int e^{-it\la/\hbar} dE_{\la}.
\eeqa
Thus, the relation
\beqa
U(t)\ket{\psi}=e^{-itH/\hbar}\ket{\psi}
\eeqa
holds for every $\ket{\psi}\in\cH_Z$.
However, those $\ket{\psi}$ satisfying 
\beqa\label{eq:6}
U(t)\ket{\psi}=\sum_{n=0}^{\infty}\frac{1}{n!}(\frac{tH}{i\hbar})^n\ket{\psi}
\eeqa
should be in the domains of all $H^n$ and satisfy
\beqa
\sum_{n=0}^{\infty}\frac{\|H^{n}\psi\|}{n!}t^{n}<\infty
\eeqa
for some $t>0$.  Such vectors $\psi$ are called analytic vectors for $H$.
Let $\ket{W}$ be an analytic vector in $\cH_W$.
From \Eq{6}, the relation
\beqa
U(t)^{\da}\ket{W}=
\sum_{n=0}^{\infty}\frac{1}{n!}(\frac{-tH}{i\hbar})^n\ket{W}
\eeqa
holds for some $t>0$.  Thus, from condition (III-A),
we can conclude that

(IV-A) $U(t)^{\da}\ket{W}\in\cH_{W}$ for some $t>0$ if $\ket{W}\in\cH_{W}$, as
long as  $\ket{W}$ is an analytic vector.

The actually derivable condition (IV-A) is much weaker than condition (IV)
claimed by WT.  In fact,
it is possible that $\cH_{W}$ may not contain a dense subset of 
analytic vectors,
and indeed this is the case for the example $H=\hat{p}$ above;
in that case, condition (IV) cannot be obtained by
manipulation of the power series \eq{4}.
Thus, WT's argument has been bogged down at condition (IV) and
never justifies their conclusion that condition (I) implies condition (II). 

Now, we can clearly see the significance of condition (I) proposed by HM
as follows.
Condition (I) means that $\cH_W$ is an invariant subspace under
$U(t)$ for all $t>0$, whereas condition (IV) means that condition (I)
implies

(I-A) $\cH_W$ is an invariant subspace under
$U(t)$ for all $-\infty<t<+\infty$.

As already stated clearly, HM required that $\cH_W$ be invariant under 
only {\em advanced} time evolution, and this is mathematically
non-equivalent to the requirement (I-A) that $\cH_W$ be invariant
under both forwards and backwards time evolution.

The failure of  (IV-A) $\Then$ (IV) is related to 
non-selfadjointness of $H$ restricted to $\cH_W$.  In fact, it is well-known
as Nelson's analytic vector theorem that the set of analytic vectors for $H$ in
$\cH_W$ is dense in $\cH_W$  if and only if $H$ restricted to $\cH_W$ is
 self-adjoint (Ref.~\cite{RS75}, p.~202).  Thus, the logical sequence
(IV-A) $\Then$ (IV)  goes through only if $H$ restricted on $\cH_W$ is
 self-adjoint, whereas in the above counter example 
it is well-known that the operator $\hat{p}$ restricted to $\cH_W$ is
symmetric but not  self-adjoint as a difficulty
in defining the momentum operator of the half-line motion.

\begin{acknowledgments}
The author thanks Masahiro Hotta, Masahiro Morikawa, Fumio Hiai, and Julio
Gea-Banacloche for useful comments and discussions.
This work was supported by the SCOPE project of the MPHPT of Japan
and by the Grant-in-Aid for Scientific Research of the JSPS.
\end{acknowledgments}

% END DOCUMENT :

\end{document}